\documentstyle[aps,prd]{revtex}
\tighten
\renewcommand{\epsilon}{\varepsilon}

\begin{document}

\title{Cosmological perturbations in the inflationary Universe}

\author{Winfried Zimdahl\footnote{Electronic address:
winfried.zimdahl@uni-konstanz.de}}
\address{Fakult\"at f\"ur Physik, Universit\"at Konstanz,
PF 5560 M678
D-78434 Konstanz, Germany}


\maketitle


\begin{abstract}

Previously defined covariant and gauge-invariant perturbation
variables,
representing, e.g., the fractional spatial energy density gradient on 
hypersurfaces of constant expansion, are used to simplify the linear 
perturbation analysis of a classical scalar field.
With the help of conserved quantities on large scales
we establish an exact first-order relation between comoving fluid
energy
density perturbations at `reentry' into the horizon and corresponding 
scalar field energy density perturbations at the first Hubble scale  
crossing
during an early de Sitter phase of a standard inflationary scenario.   

\end{abstract}

\section{Introduction}
Inflationary cosmology tries to trace back the origin of structures in 
the Universe to quantum fluctuations of a scalar field during an early 
de Sitter phase (see, e.g., \cite{MFB,LL,Boe,KoTu}).
Like all length scales the corresponding perturbation wavelengths are 
stretched out tremendously in this period and may become larger than 
the Hubble length which is constant at that time.
After the inflation has finished, i.e., when the universe has entered 
a standard Friedmann-Lema\^{\i}tre-Robertson-Walker
(FLRW) stage, the wavelength of the large-scale perturbation increases 
less than the Hubble length, resulting in an inward crossing of the 
latter
which at this time represents the particle horizon.
The first problem to tackle in this scenario is the quantum theoretical 
characterization of vacuum fluctuations of the scalar field in the de 
Sitter phase (see, e.g., \cite{MFB,LL,Boe,KoTu}).
The second one is the transition from the quantum to the classical
realm, i.e., the transformation of quantum fluctuations into classical 
energy density perturbations \cite{Lu,ChiMu}.
In this paper we are interested in the classical, general relativistic 
aspects of cosmological perturbation theory.
We will therefore neither be concerned with the quantum generation nor 
with the transition to the classical stage.
We shall assume that the perturbations behave classically once they 
have
left the "Hubble horizon".

According to the standard inflationary picture the presently observed 
large-scale inhomogeneities have been beyond the "Hubble horizon"
since the time at which the corresponding perturbation wavelengths
crossed the Hubble scale under the "slow-roll" conditions of a scalar 
field driven inflationary phase.
The large-scale fluctuations now carry an imprint of the fluctuations 
then.
While this general picture meanwhile has become common wisdom, the
establishing of a detailed link between these two far remote  
cosmological
periods is anything but obvious.
Therefore, the question of matching perturbations between cosmological 
epochs with different equations of state is of interest
(see, e.g., the recent debate in \cite{Grish,DeMu,Cald,MaSch,Goe}).  
It is the main purpose of this paper to provide a (hopefully)
transparent picture of this connection within a covariant and  
gauge-invariant
perturbation approach.
This will be achieved by using previously defined covariant and
gauge-invariant perturbation variables representing, e.g., the
fractional, spatial energy density gradient on hypersurfaces of
constant expansion, which allows us to characterize
conserved  quantities on large scales in a spatially flat universe
in a simple way.
We will find exact first-order relations between large-scale
conservation quantities and comoving energy density gradients at
far remote cosmological epochs, leading to definite expressions of
the comoving fluid energy density perturbations at "reentry" into the 
horizon in terms of corresponding energy density perturbations of
the scalar field in the inflationary phase.

Our perturbation analysis does not explicitly introduce perturbations 
of the metric tensor but is on the line of the "fluid-flow approach"  
used by Hawking \cite{Hawk}, Olson \cite{Ols}, Lyth \cite{Ly}, Lyth 
and Mukherjee \cite{LyMu}, Lyth and Stewart \cite{LyStew}.
Our basic perturbation variables are covariant and gauge-invariant
quantities of the type introduced by Ellis and Bruni \cite{EB} and
Jackson \cite{Jack}, suitably generalized in \cite{Z2CQG}.

Together with the stretching of all cosmic distances during the de
Sitter stage, a kind of smoothing effect, sometimes called
"cosmological no-hair theorem" has been conjectured
\cite{HM,FW,GP,BG}, stating that "any locally measurable
perturbation about the de Sitter metric is
damped exponentially..." (\cite{BG}).
At the first glance this seems to be contradictory to the
above-mentioned, well-known existence of conserved perturbation
quantities
on large scales \cite{BST}.
We shall clarify, which of the covariant perturbation quantities
are exponentially damped, which are conserved and how these different 
quantities are related.
Especially, we find that the (during the de Sitter phase) exponentially 
decaying perturbations cannot be neglected but start to grow again
as soon as the universe enters the standard FLRW phase.
The picture of decaying and resurrecting large-scale perturbations is 
shown to be completely equivalent to the alternative description in 
terms
of conserved quantities.

The paper is organized as follows.
In section II we introduce the basic scalar field dynamics.
Section III recalls, in a fluid picture, the cosmological perturbation 
dynamics in terms of the Ellis-Bruni-Jackson variables \cite{EB,Jack}. 
In section IV we simplify the scalar field perturbation analysis by 
rewriting the relevant equations in terms of covariant and
gauge-invariant
perturbation quantities defined with respect to hypersurfaces of
constant
expansion (constant Hubble parameter).
In section V we find the gauge-invariant and covariant connection
between
large-scale scalar field perturbations during an early de Sitter phase 
and corresponding fluid energy density perturbations at "reentering" 
the
horizon during the subsequent FLRW period.
The final section (Section VI) summarizes the conclusions of the paper.

\section{Scalar field dynamics}
The early Universe is assumed to be described by the energy momentum 
tensor $T _{mn}$ of a minimally coupled scalar field $\phi $
\begin{equation}
T _{mn} = g ^{a}_{m}g ^{b}_{n} \phi _{,a}\phi _{,b}
- g _{mn}\left(\frac{1}{2}g ^{ij}\phi _{,i}\phi _{,j}
+ V \left(\phi  \right) \right)\ .
\label{1}
\end{equation}
Assuming furthermore $\phi _{,a}$ to be timelike, one may define a
unit 4-vector
\begin{equation}
u _{i} \equiv  - \frac{\phi _{,i}}
{\sqrt{-g ^{ab}\phi _{,a}\phi _{,b}}}
\label{2}
\end{equation}
with $u ^{i}u _{i} = -1$.
It is well-known \cite{Mad} that the expression (\ref{1}) may be
rewritten
in the perfect fluid form
\begin{equation}
T _{mn} = \rho u _{m}u_{n}
+ p  h _{mn}
\label{3}
\end{equation}
with $h ^{mn} = g ^{mn} + u ^{m} u ^{n}$,
the energy density
\begin{equation}
\rho  = \frac{1}{2}\dot{\phi }^{2} + V \left(\phi  \right)\ ,
\label{4}
\end{equation}
where $\dot{\phi } \equiv  \phi _{,a}u ^{a}$,
and the pressure
\begin{equation}
p  = \frac{1}{2}\dot{\phi }^{2} - V \left(\phi  \right)\ .
\label{5}
\end{equation}
Obviously one has
\begin{equation}
\psi \equiv  \dot{\phi } = \sqrt{-g ^{ab}\phi _{,a}\phi _{,b}} \ .
\label{6}
\end{equation}

The scalar field dynamics may be written as
\begin{equation}
\dot{\rho } =
- \Theta \left(\rho
+ p  \right)
\label{7}
\end{equation}
with
\begin{equation}
\Theta \equiv u^{i}_{\ ;i}\ ,
\label{8}
\end{equation}
and
\begin{equation}
\left(\rho
+ p\right) \dot{u}_{}^{m} =
- p _{,k}h^{mk}\ ,
\label{9}
\end{equation}
where $\dot{u}^{m} \equiv u^{m}_{;n}u^{n}$.
Since the vorticity
$\omega _{ab} = h _{a}^{c}h _{b}^{d}u _{\left[c;d \right]} $ vanishes 
for a scalar field, the Raychaudhuri equation for the case of a
vanishing
cosmological constant becomes
\begin{equation}
\dot{\Theta } + \frac{1}{3}\Theta ^{2} + 2 \sigma ^{2}
- \dot{u}^{a}_{;a} + \frac{\kappa}{2}
\left(\rho + 3 p  \right) = 0 \ ,
\label{10}
\end{equation}
where
\begin{equation}
\sigma^{2} \equiv \frac{1}{2}\sigma_{ab}\sigma^{ab}\ ,
\ \ \ \ \ \ \ \
\sigma_{ab} = h_{a}^{c}h_{b}^{d}u_{\left(c;d\right)}
- \frac{1}{3}\Theta_{}h_{ ab}\ .
\label{11}
\end{equation}
The 3-curvature scalar of the hypersurfaces orthogonal to
$u ^{a}$ is
\begin{equation}
{\cal R} = 2 \left(- \frac{1}{3}\Theta ^{2}
+ \sigma ^{2}  + \kappa \rho \right)\ .
\label{12}
\end{equation}
We recall that all the relations (\ref{7}) - (\ref{12}) are valid both 
for a scalar field and for a conventional, irrotational perfect fluid. 
\section{Perturbation theory on comoving hypersurfaces}
Introducing a length scale $S$ by
\begin{equation}
\frac{1}{3}\Theta \equiv \frac{\dot{S}}{S}\ ,
\label{13}
\end{equation}
we define the comoving spatial gradient of the expansion according to
\begin{equation}
t _{a} \equiv S h ^{c}_{a} \Theta _{,c} \ .
\label{14}
\end{equation}
Following \cite{Jack}, the density inhomogeneities will be
characterized by
the quantity
\begin{equation}
D _{a} \equiv
\frac{S h ^{c}_{a}\rho _{,c}}
{\rho + p } \ ,
\label{15}
\end{equation}
being the comoving fractional spatial gradient of the energy density. 
Likewise, the pressure perturbations are described by
\begin{equation}
P _{a} \equiv
\frac{S h ^{c}_{a}p _{,c}}
{\rho  + p } \ ,
\label{16}
\end{equation}
which allows us to write Eq.(\ref{9}) as
\begin{equation}
S \dot{u}_{m} = - P _{m}\ .
\label{17}
\end{equation}
For later use we also define the 3-curvature perturbation variable
\begin{equation}
r _{a} \equiv  S h ^{c}_{a}{\cal R}_{,c}\ .
\label{18}
\end{equation}
Differentiating Eq.(\ref{7}) and projecting orthogonal to $u _{a}$
yields
\begin{equation}
h _{n}^{a} \dot{D}_{a}
+ \frac{\dot{p}}
{\rho  + p} D _{n}
+   \sigma ^{c}_{n} D _{c}
+ t _{n} = 0 \ .
\label{19}
\end{equation}
Analogously, one obtains from Eq.(\ref{10}),
\begin{eqnarray}
h _{n}^{a} \dot{t}_{a}&=&
- \dot{\Theta }P _{n}
-  \sigma ^{c}_{n} t _{c}
- \frac{2}{3} \Theta t _{n}
- 2S h ^{c}_{n} \left(\sigma ^{2}\right)_{,c}\nonumber\\
&& + S h ^{c}_{n} \left( \dot{u}^{a}_{;a}\right)_{,c} -
\frac{\kappa }{2} \left(\rho
+ p \right)
\left[D _{n} + 3 P _{n}\right]\ .
\label{20}
\end{eqnarray}
For the case of vanishing vorticity the set of equations (\ref{19}) 
and (\ref{20}) is completely general, i.e., it holds for scalar fields
and conventional perfect fluids
(cf. Eqs.(29) and (30) in \cite{Jack}).
Let us now assume the spatial gradients, i.e., the inhomogeneities, 
as well as
$\sigma $ to be small.
This is equivalent to assuming an almost homogeneous and isotropic
universe.
We regard
the quantities $D _{a}$, $P _{a}$,
$t _{a}$ and $\sigma $ as small first-order quantities on a homogeneous 
and isotropic background characterized by the zeroth-order
relations (superscript "$0$")
\begin{equation}
\kappa \rho ^{^{\left(0 \right)}}
= \frac{1}{3}\left(\Theta ^{^{\left(0 \right)}} \right)^{2}
+ \frac{1}{2}{\cal R}^{^{\left(0 \right)}} \ ,
\label{21}
\end{equation}
and
\begin{equation}
\dot{\Theta }^{^{\left(0 \right)}}
+ \frac{3}{2}\kappa \left(\rho ^{^{\left(0 \right)}}
+ p ^{^{\left(0 \right)}}\right)
= \frac{1}{2} {\cal R}^{^{\left(0 \right)}} \ ,
\label{22}
\end{equation}
with
\begin{equation}
{\cal R}^{^{\left(0 \right)}} = \frac{6 k}{a ^{2}} \ ,
\label{23}
\end{equation}
where $a$ is the scale factor of the Robertson-Walker metric.
In linear order in the inhomogeneities the acceleration term in
(\ref{20}) reduces to
\begin{equation}
S h ^{c}_{n} \left( \dot{u}^{a}_{;a}\right)_{,c} =
- \frac{\nabla ^{2}}{a ^{2}} P _{n}\ .
\label{24}
\end{equation}
Spatial differentiation of Eq.(\ref{12}) yields, in linear order,
the following relation between the first-order quantities
$r _{a}$, $t _{a}$ and
$D _{a}$
\begin{equation}
r _{a} = - \frac{4}{3}t _{a}
+ 2 \kappa \left(\rho
+ p  \right) D _{a}\ .
\label{25}
\end{equation}
(Since the coefficients of $r _{_{a}}$, $t _{_{a}}$, $D _{_{a}}$,
and $P _{_{a}}$ are always of zeroth order in a linear theory,
we omit the superscript "$0$" from now on).

Taking into account the zeroth-order relation $h _{0}^{n} = 0$, the 
system of equations (\ref{19}) and (\ref{20}) becomes, in linear
order in the spatial gradients,
\begin{equation}
\dot{D}_{\mu }
+ \frac{\dot{p}}{\rho
+ p } D _{\mu }
+ t _{\mu } = 0 \ ,
\label{26}
\end{equation}
and
\begin{equation}
\dot{t}_{\mu }  =
- \frac{2}{3} \Theta t _{\mu }
- \frac{\kappa }{2} \left(\rho
+ p \right) D _{\mu }
- \left(\frac{1}{2}{\cal R} +
\frac{\nabla ^{2}}{a ^{2}}  \right) P _{\mu }  \ .
\label{27}
\end{equation}
The system of equations (\ref{26}) and (\ref{27}) is valid for a
perfect fluid with arbitrary equation of state including a scalar
field.
In order to reformulate the dynamics for the latter case in terms
of suitable  scalar field quantities we
realize that \cite{BED}
\begin{equation}
h ^{c}_{a}\phi _{,c} = 0 \ .
\label{28}
\end{equation}
A suitable scalar field quantity to characterize spatial
inhomogeneities
is \cite{BED}
\begin{equation}
\chi _{a} \equiv \frac{Sh ^{c}_{a}\psi _{,c}}{\psi }\ .
\label{29}
\end{equation}
Because of the relations (\ref{4}), (\ref{5}), (\ref{15}),
(\ref{16}) and (\ref{28}) we have (the superscript "$s$" denotes
a scalar field quantity)
\begin{equation}
D _{a}^{\left(s \right)} = P _{a}^{\left(s \right)} = \chi _{a}\ ,
\label{30}
\end{equation}
and for a scalar field the system of equations (\ref{26})
and (\ref{27}) may be written as
\begin{equation}
\dot{\chi }_{\mu } + \left(\Theta
+ 2 \frac{\dot{\psi }}{\psi } \right) \chi _{\mu } + t _{\mu } = 0 \ ,
\label{31}
\end{equation}
and
\begin{equation}
\dot{t}_{\mu } = - \frac{2}{3} \Theta t _{\mu }
- \left[\frac{\kappa}{2}\psi ^{2} + \frac{1}{2}{\cal R}
+ \frac{\nabla ^{2}}{a ^{2}} \right] \chi _{\mu } \ ,
\label{32}
\end{equation}
respectively.
It is possible to define a formal sound velocity according to
\begin{equation}
c _{s}^{2} \equiv  \frac{\dot{p}}{\dot{\rho }}
= - \left(1 + \frac{2 \dot{\psi } }{\Theta \psi } \right)
= 1 + \frac{2 V ^{\prime } }{\Theta \psi }\ .
\label{33}
\end{equation}
Equations (\ref{31}) and (\ref{32}) may be combined to yield a
second-order differential equation for $\chi _{\mu }$:
\begin{eqnarray}
&&\ddot{\chi }_{\mu } +\left(\frac{2}{3} - c _{s}^{2} \right)
\Theta \dot{\chi }_{\mu } \nonumber\\
&&\mbox{\ }
- \left[\Theta \left(c _{s}^{2} \right)^{\displaystyle \cdot}
+ \frac{\kappa}{2}\left(\rho - 3p \right)c _{s}^{2}
+ \frac{\kappa}{2}\left(\rho + p \right)
+ \frac{{\cal R}}{2}\left(1 - c _{s}^{2} \right)
+ \frac{\nabla ^{2}}{a ^{2}}\right] \chi _{\mu } = 0 \ . \nonumber\\
\label{34}
\end{eqnarray}
For comparison we write down the corresponding equation for the
fractional energy density gradient of a perfect fluid
(superscript "$f$"), $D _{\mu }^{\left(f \right)}$:
\begin{eqnarray}
\ddot{D}_{\mu }^{^{\left(f \right)}} &+& \left(\frac{2}{3}
- v _{s}^{2} \right)
\Theta \dot{D}_{\mu }^{^{\left(f \right)}} \nonumber\\
&-& \left[\Theta \left(v _{s}^{2} \right)^{\displaystyle \cdot}
+ \frac{\kappa}{2}\left(\rho - 3p \right)v _{s}^{2}
+ \frac{\kappa}{2}\left(\rho + p \right)
+ v _{s}^{2}\frac{\nabla ^{2}}{a ^{2}}\right]
D _{\mu }^{^{\left(f \right)}} = 0 \ .
\label{35}
\end{eqnarray}
The quantity
$v _{s} $ is the adiabatic sound velocity of the fluid, determined by 
$v_{s}^{2} = \left(\partial p/ \partial \rho  \right)_{adiabatic}$. 
The equations (\ref{34}) and (\ref{35}) are very similar.
The differences are due to the different relations between the
pressure perturbations and the energy density perturbations in both 
cases.
While according to (\ref{30}) the quantities $D _{a}$ and $P _{a}$
coincide in the case of a scalar field the corresponding relation
for a fluid is
\cite{Z2CQG}
\begin{equation}
P _{a}^{^{\left(f \right)}} = v_{s}^{2} D _{a}^{^{\left(f \right)}}\ .
\label{36}
\end{equation}

In terms of the potential and its derivatives, equation (\ref{34})
becomes
\begin{eqnarray}
\ddot{\chi }_{\mu } &-& \frac{1}{3}
\left(1 + \frac{6 V ^{\prime } }{\Theta \psi } \right)
\Theta \dot{\chi }_{\mu } \nonumber\\
&-& \left[2 V ^{\prime \prime } + 2 \kappa V
+ \frac{2 V ^{\prime }}{\Theta \psi }
\left(5 + \frac{3 V ^{\prime }}{\Theta \psi } \right)
\frac{\Theta ^{2}}{3}
+ \frac{\nabla ^{2}}{a ^{2}}\right] \chi _{\mu } = 0 \ .
\label{37}
\end{eqnarray}
\section{Linear perturbation theory on hypersurfaces of constant
expansion}
All the first-order quantities $D _{a}$,
$P _{a}$, $\chi _{a}$, $t _{a}$,
$r _{a}$, have physical interpretations on comoving hypersurfaces.
It was shown in \cite{Z2CQG} that the perfect fluid dynamics
simplifies if written in terms of covariant and gauge-invariant
variables with physical interpretations on hypersurfaces of
constant curvature, constant expansion, or constant energy density. 
These variables correspond to certain combinations of
the Ellis-Bruni-Jackson variables. E.g., the quantity
\begin{equation}
D _{a}^{\left(ce \right)} \equiv
D _{a} - \frac{\dot{\rho }}{\rho + p}
\frac{t _{a}}{\dot{\Theta }}
\label{38}
\end{equation}
represents in first order
the fractional, spatial gradient of the energy density on hypersurfaces 
of constant expansion (superscript "ce").
This interpretation becomes obvious if one writes down the first-order 
expressions for $h ^{c}_{a}\rho _{,c}$ and
$h ^{c}_{a}\Theta  _{,c}$ occuring in
$D _{a}$ and $t _{a}$ (see the definitions (\ref{15}) and (\ref{14})), 
respectively:
\begin{equation}
\left(h ^{c}_{0}\rho _{, c} \right)^{\hat{}} = 0\ ,\ \ \ \ \ \ \
\left(h ^{c}_{\alpha}\rho _{, c} \right)^{\hat{}}
= \hat{\rho }_{,\alpha} + \hat{u}_{\alpha}
\dot{\rho }^{^{\left(0 \right)}}\ ,
\label{39}
\end{equation}
and
\begin{equation}
\left(h ^{c}_{0}\Theta  _{, c} \right)^{\hat{}} = 0\ ,\ \ \ \ \ \ \ 
\left(h ^{c}_{\alpha}\Theta  _{, c} \right)^{\hat{}}
= \hat{\Theta  }_{,\alpha}
+ \hat{u}_{\alpha}\dot{\Theta }^{^{\left(0 \right)}}\ ,
\label{40}
\end{equation}
where the superscripts "$0$" again denote the homogeneous and
isotropic zeroth
order while the quantities with carets are of first order in the
deviations from homogeneity and isotropy.
Using the relations (\ref{14}), (\ref{15}), (\ref{39}), and (\ref{40}) 
in Eq.(\ref{38}),
the quantity $D _{a}^{\left(ce \right)}$ becomes, in first order,
\begin{equation}
\hat{D}_{a}^{\left(ce \right)}
= \frac{S \hat{\rho }_{,\mu }}
{\rho ^{^{\left(0 \right)}} + p ^{^{\left(0 \right)}}}
- \frac{\dot{\rho }^{^{\left(0 \right)}}}
{\dot{\Theta }^{^{\left(0 \right)}}}
\frac{S \hat{\Theta  }_{,\mu }}
{\rho ^{^{\left(0 \right)}} + p ^{^{\left(0 \right)}}}\ .
\label{41}
\end{equation}
While $D _{a}^{\left(ce \right)}$, defind in Eq.(\ref{38}),
is an exact, covariant quantity without any reference to  
perturbation theory,
the perturbation quantity $\hat{D} _{a}^{\left(ce \right)}$ is  
gauge-invariant
by construction.
Since $\hat{D}_{a}^{\left(ce \right)}$ is gauge-invariant,
the combination on the right-hand side (RHS) of Eq.(\ref{41})
is gauge-invariant as well, while each term on its own is not.
Therefore, in a gauge with $\hat{\Theta }_{,\mu } = 0$
(notice that $\Theta$ itself is only determined up to a constant), i.e., 
vanishing perturbations of the expansion, the covariant and
gauge-invariant quantity
$D_{\mu}^{\left(ce \right)}$ coincides in first order with the
fractional density gradient $S \hat{\rho }_{,\mu }/
(\rho ^{^{\left(0\right)}}
+ p ^{^{\left(0 \right)}})$, where $\rho$ according to the
decomposition
(\ref{3}) is the energy density measured by a comoving observer.
A different observer, moving, e.g., with a 4- velocity $n^{a}$,
normal to hypersurfaces
$\hat{\Theta }_{,\mu } = 0$, would interpret a different quantity,
namely
$\mu = T_{ab}n^{a}n^{b}$ as energy density. (Different from the
observer moving with
$u^{a}$ he would also measure an heat flux.)
The densities $\rho $ and $\mu $
are related by (see \cite{KiEll,BDE})
\[
\mu = \rho \cosh ^{2}\beta + p \sinh ^{2}\beta \ ,
\]
where $\beta (t)$ is the "hyperbolic angle of tilt" given by $\cosh 
\beta = -
u ^{a}n _{a}$. It follows that the perturbation quantity $\hat{\rho 
}$ for a comoving (with $u ^{a}$) observer will generally not coincide 
with
the perturbation $\hat{\mu } $, i.e., $\hat{\rho }$ will not coincide 
with
the energy density perturbation on hypersurfaces of constant expansion. 
In the present case, however, $u^{a}$
and $n ^{a}$ coincide in zeroth order and $\beta$ may be considered 
as small ($\beta \ll 1$).
Under this condition the difference between $\mu $ and $\rho $
is of second order in $\beta $. Consequently, in linear perturbation
theory we may identify the quantities $\hat{\mu }$ and $\hat{\rho }$. 
It follows that the quantity (\ref{41}) may be interpreted as the  
fractional
energy density gradient on constant expansion hypersurfaces in a
similar sense
in which $D_{a}$ is interpreted as fractional energy density
gradient on comoving hypersurfaces.
Similar statements hold for the other perturbation variables.

A curvature variable on hypersurfaces of constant expansion,
$r _{a}^{\left(ce \right)}$, is given by
\begin{equation}
r _{a}^{\left(ce \right)} \equiv  r _{a}
- \dot{{\cal R}}\frac{t _{a}}{\dot{\Theta }}\ .
\label{42}
\end{equation}
For vanishing background curvature ${\cal R} = 0$ the latter
quantity coincides with $r _{a}$, i.e.,
$r _{a}^{\left(ce \right)}
\stackrel{\left(k = 0 \right)}{=} r _{a}$.
In terms of the variables $D _{a}^{\left(ce \right)}$ and
$r _{a}^{\left(ce \right)}$ the relation (\ref{25})
reduces to \cite{Z2CQG}
(cf.\cite{H1})
\begin{equation}
r_{a}^{\left(ce \right)}
= 2 \kappa \left(\rho + p \right) D _{a}^{\left(ce \right)} \ .
\label{43}
\end{equation}
An expansion perturbation variable does no longer occur.
Effectively, the number of variables has been reduced \cite{Z2CQG}.

Analogously, for universes with nonvanishing background
curvature one may define the fractional energy density on constant
curvature hypersurfaces
(superscript "cc")
\begin{equation}
D _{a}^{\left(cc \right)} \equiv
D _{a} - \frac{\dot{\rho }}{\rho + p}
\frac{r _{a}}{\dot{{\cal R} }}\ .
\label{44}
\end{equation}
The corresponding perturbation variable of the expansion is
\begin{equation}
t _{a}^{\left(cc \right)} \equiv  t _{a}
- \dot{\Theta }\frac{ r_{a}}{\dot{{\cal R} }}\ .
\label{45}
\end{equation}
In terms of the variables (\ref{44}) and (\ref{45}), defined with
respect to constant curvature hypersurfaces, the relation (\ref{25}) 
between the three variables
$r _{a}$, $t _{a}$, $D _{a}$ becomes \cite{Z2CQG} (cf.\cite{H1})
\begin{equation}
\frac{4}{3}\Theta t _{a}^{\left(cc \right)}
= 2 \kappa \left(\rho + p \right) D _{a}^{\left(cc \right)}\ .
\label{46}
\end{equation}
No curvature perturbation variable does occur in Eq.(\ref{46}).
The use of either $D _{a}^{\left(ce \right)}$ or
$D _{a}^{\left(cc \right)}$ as basic variable simplifies the
perturbation dynamics significantly \cite{Z2CQG}.
Since for universes with zero background curvature the variable
$D _{a}^{\left(cc \right)}$ is not defined we will use
$D _{a}^{\left(ce \right)}$ as independent variable in terms of which 
the system of equations (\ref{26}) and (\ref{27}) may be written in 
the form
\begin{equation}
\left[a ^{2}\dot{\Theta }
D _{\mu }^{\left(ce \right)}\right]^{\displaystyle \cdot}
=  3 k \Theta \left[\frac{\dot{p}}{\dot{\rho }} D _{\mu}
- P _{\mu}\right]
- a ^{2}\Theta \frac{\nabla ^{2}}{a ^{2}}P _{\mu } \ .
\label{47}
\end{equation}
This relation is equivalent to Eqs.(\ref{26}) and (\ref{27}).
It contains the entire linear perturbation dynamics and holds both
for a scalar field, characterized by the relations (\ref{30}),
and for a fluid.
For a fluid equation of state $p = p \left(\rho  \right)$ one has
\begin{equation}
\dot{p} = v _{s}^{2} \dot{\rho }\ ,
\label{48}
\end{equation}
and it follows with Eq.(\ref{36}) that the bracket in the
first term on the RHS
of (\ref{47}) vanishes.

For $D _{\mu } = D _{\left(n \right)}\nabla _{\mu }
Q _{\left(n \right)}$,
$P _{\mu}
= P _{\left(n \right)}
\nabla _{\mu }Q _{\left(n \right)}$ and
$D _{\mu }^{\left(ce\right)}
= D _{\left(n \right)}^{\left(ce\right)} \nabla _{\mu }Q _{\left(n  
\right)}$
(and corresponding relations for the other perturbation quantities),  
where
the $Q _{\left(n \right)}$ satisfy the Helmholtz equation $\nabla ^{2}Q
_{\left(n \right)} = - n ^{2} Q _{\left(n \right)}$ one has (\cite
{LiKha,Harr,KoSa,DB}) $n ^{2} = \nu ^{2}$ for $k = 0$ and $n ^{2} =  
\nu ^{2}
+ 1$ for $k = -1$, where $\nu$ is continous and related to the physical
wavelength by $\lambda = 2 \pi a/\nu$. For $k = +1$ the eigenvalue
spectrum
is discrete, namely $n ^{2} = m \left(m + 2 \right)$ with
$m = 1,2,3....$.

It follows that for $k = 0$ the spatial gradient terms  on
the RHS of Eq.(\ref{47})
may be neglected on large perturbation scales ($\nu \ll 1$)
and the quantity
$a ^{2}\dot{\Theta }
D _{\left(\nu\right) }^{\left(ce \right)}$ is a conserved quantity  
both
for a perfect fluid and a scalar field.

In the scalar field case the curvature term on the RHS of Eq.(\ref{47}) 
becomes
\begin{equation}
3 k \Theta \left[\frac{\dot{p}}{\dot{\rho }} D _{a}^{\left(s \right)} 
- P _{a}^{\left(s \right)}\right] =
- 6 k \left(\Theta + \frac{\dot{\psi }}{\psi } \right) \chi
= - 6 k \frac{V ^{\prime }}{\psi }\chi \ .
\label{49}
\end{equation}
Using the relations (\ref{38}), (\ref{30}),
(\ref{31}) and (\ref{32}), we find
\begin{equation}
a ^{2}\dot{\Theta }
D _{\mu }^{\left(ce \right)} = a ^{2}
\left[\left(\dot{\Theta } - \Theta
\left(\Theta + 2 \frac{\dot{\psi }}{\psi } \right)\right)
\chi _{\mu }  - \Theta \dot{\chi }_{\mu }\right]\ .
\label{50}
\end{equation}
Inserting Eqs.(\ref{49}) and (\ref{50}) into Eq.(\ref{47}) one gets
\begin{equation}
\left\{a ^{2}\left[\left(\dot{\Theta } + \Theta ^{2}c _{s}^{2} \right)
\chi _{\mu } - \Theta \dot{\chi }_{\mu } \right]\right\}
^{\displaystyle \cdot}
= \left[6 k \frac{V ^{\prime }}{\psi }
- \Theta \nabla ^{2}\right]\chi _{\mu }\ ,
\label{51}
\end{equation}
which is a different, more compact way of writing Eqs.(\ref{34})
or (\ref{37}).

It is obvious from Eqs.(\ref{43}) and (\ref{22}) that for $k = 0$
the relation
\begin{equation}
r _{a}^{\left(ce \right)}
\stackrel{\left(k = 0 \right)}{=} r _{a}
= \frac{3}{4}\left[\Theta \dot{\chi }_{\mu }
- \left(\dot{\Theta } - \Theta
\left(\Theta + 2 \frac{\dot{\psi }}{\psi } \right) \right)
\chi _{\mu }\right]
\label{52}
\end{equation}
is valid.
The large-scale conservation quantity (for a flat background)
$a ^{2}r _{a}$ coincides with the quantity $C _{a}$ used by Dunsby
and Bruni \cite{DB}.

\section{Scalar field perturbations in the de Sitter phase and their 
relation to fluid perturbations in the FLRW epoch}
Let us now apply the equations (\ref{31}) and (\ref{32}), or,
equivalently Eq.(\ref{51}) for the general scalar field dynamics
to the "slow-roll" period of an inflationary universe
with $k = 0$.
The "slow-roll" conditions are
\begin{equation}
\dot{\psi } \ll \Theta \psi \ ,\ \ \ \ \ \ \ \
\Theta \approx const \ .
\label{57}
\end{equation}
The relation (\ref{50}) specifies to
\begin{equation}
a ^{2}\dot{\Theta }
D _{\mu }^{\left(ce \right)}
= - a ^{2}\Theta \left[\dot{\chi }_{\mu }
+ \Theta \chi _{\mu }\right]
= - \frac{\Theta }{a}
\left[a ^{3} \chi _{\mu } \right]^{\displaystyle \cdot}\ .
\label{58}
\end{equation}
>From Eq.(\ref{47}) and the discussion below Eq.(\ref{48})
we know that this quantity is conserved on
large scales for $k = 0$.
Denoting the conserved quantity by $- E _{\left(\nu\right) }$, i.e., 
\begin{equation}
- E _{\left(\nu\right) } \equiv   a ^{2}\dot{\Theta }
D _{\left(\nu\right) }^{\left(ce \right)}\ , \ \ \ \ \ \ \ \ \ \
\left(\nu \ll 1\right)
\label{59}
\end{equation}
equation (\ref{58}) becomes
\begin{equation}
\left[\chi _{\left(\nu\right)}a ^{3} \right]^{\displaystyle \cdot}
= \frac{a}{\Theta }E _{\left(\nu\right) }\, \ \ \ \ \ \ \ \ \
\left(\nu \ll 1\right)
\label{60}
\end{equation}
on large scales.
Taking into account $a \sim \exp{\left(H t\right)}$
with $3 H \equiv  \Theta $,
the solution of Eq.(\ref{60}) is
\begin{equation}
\chi _{\left(\nu\right) } = \frac{3}{\Theta ^{2}}
\frac{E _{\left(\nu\right)}}
{a ^{2}}
+ \frac{C _{\left(\nu\right) }^{\left(s \right)}}{a ^{3}}\
,\ \ \ \ \ \ \ \ \ \ \ \ \left(\nu \ll 1\right)
\label{61}
\end{equation}
where $C _{\left(\nu\right) }^{\left(s \right)}$ is an integration
constant
for the scalar field case (superscript "s").
The large-scale solution (\ref{61}) is also found from the second-order 
equation (\ref{37}) for $\chi _{\mu }$, which
under the "slow-roll" conditions
\begin{equation}
\frac{V ^{\prime }}{\Theta \psi } \approx - 1 \ ,\ \ \ \ \ \ \
V ^{\prime \prime } \ll \frac{1}{3}\Theta ^{2}\ ,
\label{62}
\end{equation}
reduces to
\begin{equation}
\ddot{\chi }_{\left(\nu\right) } + \frac{5}{3}\Theta \dot{\chi  
}_{\left(\nu\right) }
- \left[{\cal R} - \frac{2}{3}\Theta ^{2}
+ \frac{\nabla ^{2}}{a ^{2}} \right] \chi _{\left(\nu\right) } = 0
\ .
\label{63}
\end{equation}

The solution (\ref{61}) describes the exponential damping of the
comoving, fractional energy density gradient on large scales
during the de Sitter phase.
It demonstrates the stability of the latter with respect to linear
perturbations and seems to
represent a gauge-invariant formulation of the cosmic
"no-hair" theorem \cite{HM,FW,GP,BG}.
While the perturbation wavelengths are stretched out tremendously, the 
amplitude of the comoving, fractional energy density gradient becomes 
exponentially small.
As will be shown below, it is {\it not} justified, however, to neglect 
these perturbations since they will resurrect during the
subsequent FLRW phase.
While the "hair" gets extremely shortend during the
de Sitter period, it is not completely extinguished but
slowly grows again afterwards.

It is of essential interest for any inflationary scenario to
establish a link between the large-scale perturbations at
Hubble scale crossing within an early de Sitter stage on the one
hand side, and at "reentering" the horizon during the subsequent
FLRW phase on the other side.
It is well known that the existence of conservation quantities is
helpful to find this connection \cite{MFB,LL,Ly,BST,H2,Hscal}.
As we shall demonstrate here the description in terms of
gauge-invariant and covariant perturbation variables provides us
with a definite expression for the large-scale comoving,
fractional energy density gradient at "reentering" the
horizon in terms of the corresponding quantity at Hubble
scale crossing during the "slow-roll" period of an
inflationary universe.

Let us assume the evolution of the cosmic medium to encompass
an early phase during which it is adequately described by a
scalar field that for some time admits the fulfillment
of the "slow-roll" conditions (\ref{57})
or (\ref{62}).
According to the standard picture (see, e.g., \cite{KoTu})
the scalar field finally decays and the universe enters the
familiar FLRW stage which initially is radiation dominated.
Since both the early scalar field dominated period and the
final FLRW phase are accessible to  (perfect) fluid
descriptions with different equations of state, a unified,
one-component picture of the cosmological evolution
including an inflationary phase is possible.
The different equations of state give rise to different values of
the sound velocity.
The time dependence of the latter is, however, fully taken into
account in our basic equation (\ref{47}).
The set of equations (\ref{26}) and (\ref{27}) or, equivalently,
Eq.(\ref{47}), is valid both for scalar fields and for fluids.

For $k = 0$ the covariant quantity
$a ^{2}\dot{\Theta }D _{\mu}^{\left(ce \right)}$ is conserved on
large scales independently of the equations of state.
Especially, the large-scale conservation of $E _{\mu }$ holds for a 
minimally coupled scalar field and for a conventional perfect fluid. 
Provided, nonadiabatic pressure perturbations during reheating do not 
substantially alter the present picture, on large scales
$E _{\mu }$ remains the {\it same} quantity
under the change from
the early scalar field dominated stage to the latter FLRW epoch.
Since we have established the connection between $E _{\mu }$ and
the other covariant and gauge-invariant perturbation quantities,
especially the comoving, fractional energy density gradients
both for scalar fields,
$D _{\mu }^{\left(s \right)} \equiv  \chi _{\mu }$, and for
conventional adiabatic fluids, $D _{\mu }^{^{\left(f \right)}}$,
we are able to express the comoving, large-scale perturbation
amplitude at the time of "reentry", say, today in terms of the
comoving, large-scale perturbation amplitude at the time of
Hubble scale crossing during the inflationary period.

Denoting the initial Hubble scale crossing time by $t _{i}$ and
the time of "reentering" the horizon during the perfect fluid
FLRW phase
by $t _{e}$, conservation of the quantity (\ref{59}) means
\begin{equation}
a ^{2}\left(t _{i} \right)
\dot{\Theta }\left(t _{i} \right)
D _{\left(\nu\right) }^{\left(ce \right)}\left(t _{i} \right)
\equiv  - E _{\left(\nu\right) }\left(t _{i} \right)
= - E _{\left(\nu\right) }\left(t _{e} \right)
\equiv  a ^{2}\left(t _{e} \right)
\dot{\Theta }\left(t _{e} \right)
D _{\left(\nu\right) }^{\left(ce \right)}\left(t _{e} \right)
\ ,\ \ \ \ \ \ \ \ \ \left(\nu \ll 1\right)\ .
\label{64}
\end{equation}
In order to relate $D _{\left(\nu\right) }\left(t _{i} \right)$ to
$D _{\left(\nu\right) }\left(t _{e} \right)$ we have first to find  
the connection
between $E _{\left(\nu\right) }\left(t _{e} \right)$ and
$D _{\left(\nu\right) }\left(t _{e} \right)$.

Using in Eq.(\ref{59}) the relations (\ref{38}),
(\ref{26}) and (\ref{27}) for $k = 0$, we obtain
\begin{equation}
E _{\left(\nu\right) } = - a ^{2}\dot{\Theta }
D _{\left(\nu\right) }^{\left(ce \right)}
= a ^{2}\left[\Theta \dot{D}_{\left(\nu\right) }^{^{\left(f \right)}} 
+ \left(\frac{\gamma }{2} - v _{s}^{2} \right)
\Theta ^{2}D _{\left(\nu\right) }^{^{\left(f \right)}} \right]\ ,
\ \ \ \ \ \ \ \left(\nu \ll 1\right)
\label{65}
\end{equation}
where $\gamma = (\rho ^{^{\left(f \right)}}
+ p ^{^{\left(f \right)}})/ \rho ^{^{\left(f \right)}}$.
The modes of interest are expected to "reenter" the the horizon at
a period with constant values of $\gamma $ and $v _{s}^{2} $.
Under such circumstances we may write
\begin{equation}
E _{\left(\nu\right) } = \Theta a ^{2}a ^{-3 \left(\gamma /2
- v _{s}^{2} \right)}
\left[a ^{3 \left(\gamma /2 - v _{s}^{2}    \right)}
D _{\left(\nu\right) }^{^{\left(f \right)}}
\right]^{\displaystyle \cdot}
,\ \ \ \ \ \ \ \ \left(\nu \ll 1\right)\ .
\label{66}
\end{equation}
Let us first consider the case that at "reentering" the universe
is still radiation dominated, i.e., $p = \rho /3$, $\gamma = 4/3$,
$v_{s}^{2} = 1/3$. In this case we have (the superscript "r"
stands for radiation)
\begin{equation}
\left[a D _{\left(\nu\right) }^{\left(r \right)}  
\right]^{\displaystyle \cdot}
= \frac{E _{\left(\nu\right) }}{\Theta a}
,\ \ \ \ \ \ \ \ \ \ \ \ \ \left(\nu \ll 1\right)\ .
\label{67}
\end{equation}
With $\Theta = 3 \dot{a}/a = 3H$
and $a = a _{0}\left(t/t _{0} \right)^{1/2}$,
integration of Eq.(\ref{67}) yields
\begin{equation}
D _{\left(\nu\right)}^{\left(r \right)}\left(t \right)
= D _{\left(\nu\right)}^{\left(r,g \right)}\left(t \right)
+ \frac{C _{\left(\nu\right) }^{\left(r \right)}}{a}\ ,
\ \ \ \ \ \ \ \ \ \ \ \ \left(\nu \ll 1\right)
\label{68}
\end{equation}
where
\begin{equation}
D _{\left(\nu\right)}^{\left(r,g \right)}\left(t \right)
= \frac{1}{9}\frac{E _{\left(\nu\right)}}{H^ 2 a^2}
\ ,
\ \ \ \ \ \ \ \ \ \ \ \ \left(\nu \ll 1\right)
\label{69}
\end{equation}
is the growing mode (superscript "$g$") of the solution (\ref{68}). 
The quantity
$C _{\left(\nu\right) }^{\left(r \right)}$ is an integration constant 
for
the radiation case.
Because of $H^ {2} a^{2} \propto a ^{-2}$ we have reproduced the  
well-known
result for the growing
( $\sim a ^{2}$) and decaying ( $\sim a ^{-1}$) perturbation
modes in a radiation dominated universe.
The advantage of the calculation via Eq.(\ref{67}) instead of
solving
a second-order differential equation for $D _{\left(\nu\right) }$,
as is
usually done, is that one obtains an explicit relation
between the conserved
quantity $E _{\left(\nu\right) }$ and the growing mode.
{\it The conserved quantity $E _{\left(\nu\right) }$ is
coupled to the growing mode only.}
Realizing that $\left(Ha\right)^{-1}$ is the comoving horizon scale 
in a radiation dominated universe,
Equation (\ref{69}) implies that the dynamics of the growing mode
$D _{\left(\nu\right)}^{\left(r,g \right)}$
is entirely determined by the square of the comoving horizon distance.  
It follows, that (up to a numerical constant) the conserved quantity 
$E _{\left(\nu\right) }$ represents the ratio of the growing part
of the comoving, fractional density perturbations to the square
of the comoving particle horizon.

Along the same lines one finds a corresponding relation for the
scalar field in the "slow-roll" phase.
Recalling $\chi _{_{\left(\nu\right) }} \equiv
D _{_{\left(\nu\right) }}^{\left(s \right)} $
from Eq.(\ref{30}),
Equation (\ref{61}) may be written as
\begin{equation}
D _{\left(\nu\right) }^{\left(s \right)} =
D _{\left(\nu\right) }^{\left(s,d \right)}
+ \frac{C _{\left(\nu\right) }^{\left(s \right)}}{a ^{3}}\ ,
\ \ \ \ \ \ \ \ \ \ \ \ \left(\nu \ll 1\right)
\ .
\label{72}
\end{equation}
$D _{\left(\nu\right) }^{\left(s,d \right)}$ is
the "dominating" mode (there is no growing mode in the present case) 
\begin{equation}
D _{\left(\nu\right)}^{\left(s,d \right)}\left(t \right)
= \frac{1}{3}\frac{E _{\left(\nu\right)}}{H^ 2 a^2}
\ ,
\ \ \ \ \ \ \ \ \ \ \ \left(\nu \ll 1\right)
\label{73}
\end{equation}
during the "slow-roll" period.
{\it $E _{\left(\nu\right) }$ couples only to
the "dominating" mode during an
early de Sitter phase.}
In the de Sitter period the quantity $\left(Ha\right)^{-1}$
represents the comoving event horizon, the square of which is
proportional
to the dominating mode
$D _{\left(\nu\right)}^{\left(s,d \right)}$.
In this era the constant quantity $E _{\left(\nu\right) }$
determines the ratio of the "dominating" part of the comoving,
fractional energy density perturbations to the comoving Hubble
horizon
for $\nu \ll 1$.

The relation
$E _{\left(\nu\right) }\left(t _{i} \right)
= E _{\left(\nu\right) }\left(t _{e} \right)$ (cf.Eq.(\ref{64}))
provides us with
\begin{equation}
D_{\left(\nu\right) }^{\left(s,d \right)}\left(t _{i} \right)
H ^{2}\left(t _{i}\right) a ^{2}\left(t _{i} \right)
= \frac{1}{3}
D _{\left(\nu\right) }^{\left(r, g \right)}\left(t _{e} \right)
H ^{2}\left(t _{e}\right) a ^{2}\left(t _{} \right)\ ,
\ \ \ \ \ \ \ \ \ \ \ \left(\nu \ll 1\right)\ .
\label{75}
\end{equation}
Since the horizon crossing conditions for a perturbation of the
constant, comoving wavelength $\lambda /a$ are
$\lambda /a \approx
\left[H\left(t _{i}\right) a\left(t _{i} \right)\right]^{-1}
\approx
\left[H\left(t _{e}\right) a\left(t _{e} \right)\right]^{-1}$,
Eq. (\ref{75}) reduces to
\begin{equation}
D_{\left(\nu\right) }^{\left(r,g\right)}\left(t _{e} \right)
\approx \frac{1}{3}
D _{\left(\nu\right) }^{\left(s,d\right)}\left(t _{i} \right)
\ \ \ \ \left(\nu \ll 1\right)\ .
\label{76}
\end{equation}
At reentry into the horizon, the comoving, fractional
energy density perturbations are about $1/3$ the
comoving, fractional
energy density perturbations at Hubble horizon crossing
during the de Sitter phase.
With other words, the corresponding "transfer function" is
$1/3$.

If the universe is already matter dominated at reentry into the
horizon,
Eq.(\ref{66}) becomes (the superscript "$m$" stands for matter)
\begin{equation}
\left[a ^{\frac{3}{2}}
D _{\left(\nu\right)}^{\left(m \right)}\right]^{\displaystyle \cdot} 
= \frac{E _{\left(\nu\right) }}{\Theta a ^{1/2}}\ ,
\ \ \ \ \ \ \ \ \ \left(\nu \ll 1\right)
\ .
\label{76a}
\end{equation}
Because of $a = a _{0}\left(t/t _{0} \right)^{2/3}$ we find
\begin{equation}
D _{\left(\nu\right) }^{\left(m \right)} =
D _{\left(\nu\right) }^{\left(m, g \right)}
+ \frac{C _{\left(\nu\right) }^{\left(m \right)}}{a ^{3/2}}
\ \ \ \ \left(\nu \ll 1\right)\ ,
\label{77}
\end{equation}
where the growing mode $D _{\left(\nu\right) }^{\left(m, g \right)}$ 
is given
through
\begin{equation}
D _{\left(\nu\right)}^{\left(m, g \right)}
= \frac{2}{15}\frac{E _{\left(\nu\right)}}{H^ 2 a^2}
\ ,
\ \ \ \ \ \ \ \ \ \ \left(\nu \ll 1\right)\ .
\label{78}
\end{equation}
The horizon crossing conditions in this case are
$\lambda /a \approx
\left[H\left(t _{i}\right) a\left(t _{i} \right)\right]^{-1}
\approx
2\left[H\left(t _{e}\right) a\left(t _{e} \right)\right]^{-1}$, which 
yields
\begin{equation}
D_{\left(\nu\right) }^{\left(m,g\right)}\left(t _{e} \right)
\approx \frac{1}{10}
D _{\left(\nu\right) }^{\left(s,d\right)}\left(t _{i} \right)
\ \ \ \ \ \ \ \ \ \ \left(\nu \ll 1\right)\ ,
\label{79}
\end{equation}
instead of Eq.(\ref{76}).
The comoving, fractional energy density perturbations
at reentry into the horizon during the matter dominated FLRW period 
are smaller than the corresponding perturbations
at Hubble horizon crossing during the "slow roll" phase
by one order of magnitude, i.e., the "transfer function" is
$1/10$.

It becomes
clear now, in which sense the exponentially decaying perturbations
$D _{\left(\nu\right) }^{\left(s,d \right)}$ (cf. Eq.(\ref{73}))
resurrect in order to yield the
fluid energy density perturbations at reentry into the
Hubble radius at $t = t _{e}$.

Let us denote by $t _{f}$ the time at which the
de Sitter phase finishes, i.e.,
$t _{i} < t _{f} < t _{e}$.
During the time interval $t _{f} - t _{i}$ the scale factor increases 
exponentially with $H = const$.
Because of the $a ^{-2}$ dependence of the "dominating" mode
(\ref{73}), the fractional, comoving
scalar field energy density perturbations are
exponentially suppressed by
many orders of magnitude, i.e.,
$D _{\left(\nu\right) }^{\left(s,d \right)}
\left(t _{f} \right)$  is vanishingly small
compared with
$D _{\left(\nu\right) }^{\left(s,d \right)}\left(t _{i} \right)$.
For $t > t _{f}$ the overall energy density perturbations are
represented
by the fluid quantity
$D _{\left(\nu\right) }^{\left(r,g \right)}$.
Assuming an instantaneous transition to the radiation dominated
FLRW stage, i.e.,
$D _{\left(\nu\right) }^{\left(s,d \right)}\left(t _{f} \right)
\approx
D _{\left(\nu\right) }^{\left(r,g \right)}\left(t _{f} \right)$,
we have $a \propto t ^{1/2}$ for $t > t _{f}$ and a
power law
growth $\propto a ^{2} \propto t$ of the quantity
$D _{\left(\nu\right) }^{\left(r,g \right)}$.
Consequently, it needs a large time interval $t _{e} - t _{f}$
during which the power law growth of
$D _{\left(\nu\right) }^{\left(r,g \right)}$ compensates
the initial exponential decay of
$D _{\left(\nu\right) }^{\left(s,d \right)}$ during the interval
$t _{f} - t _{i}$.
Finally, at $t = t _{e}$, i.e., at the time at which the
perturbation crosses the horizon inwards, the
fluid energy density perturbations
$D _{\left(\nu\right) }^{\left(r,g \right)}$ or
$D _{\left(\nu\right) }^{\left(m,g \right)}$ are
almost of the same order again as the scalar field perturbations
$D _{\left(\nu\right) }^{\left(s,d\right)}$ at the first horizon
crossing time $t _{i}$.

Obviously, the description in terms of comoving energy density
perturbations that initially become exponentially small but
afterwards start to grow again is completely equivalent to the
characterization with the help of conserved quantities.
It is the advantage of our covariant approach that it is capable of 
providing us with a clear and transparent picture of the different
aspects of the behaviour of large-scale perturbations in an
inflationary Universe.

\section{Summary}
Within a covariant and gauge-invariant approach we have investigated 
first-order perturbations in the inflationary universe.
Perturbation quantities at the time of horizon entry during the
FLRW phase were related to corresponding quantities at the Hubble
scale crossing time in an early de Sitter phase.
The perturbation dynamics was simplified by using a covariant,
large-scale conservation quantity, expressed in terms of energy
density perturbations on hypersurfaces of constant expansion.
The relations between this conserved quantity and covariantly
defined,
comoving, fractional energy density perturbations were clarified
for different periods of the cosmological evolution.
The conservation quantity turned out to be uniquely linked to the
dominating comoving perturbation modes both in the "slow-roll"
period with exponential expansion and in the FLRW stage.
Equivalent to the characterization in terms of conserved quantities, 
the large-scale dynamics may be described with the help of
comoving, fractional energy density perturbations.
The latter are exponentially suppressed during the de Sitter
phase but resurrect as soon as the universe enters the FLRW
period.
The corresponding "transfer functions" are $1/3$ in the
radiation dominated FLRW phase and $1/10$ in the dust era.

\acknowledgments
This paper was supported by the Deutsche Forschungsgemeinschaft.
I thank the referee for useful suggestions.

\end{document}